\documentclass[twocolumn,aps,superscriptaddress,showpacs]{revtex4}
\usepackage{amssymb}
\usepackage{amsmath}
\usepackage{graphicx}

\begin{document}

\title{Initialization effect in heavy-ion collisions at intermediate energies}
\author{Gao-Chan Yong}\affiliation{Institute of Modern Physics, Chinese Academy of
Sciences, Lanzhou 730000, China}
\author{Yuan Gao}
\affiliation{School of Information Engineering, Hangzhou Dianzi
University, Hangzhou 310018, China}
\author{Wei Zuo}
\affiliation{Institute of Modern Physics, Chinese Academy of
Sciences, Lanzhou 730000, China}
\author{ Xun-Chao Zhang}
\affiliation{Institute of Modern Physics, Chinese Academy of
Sciences, Lanzhou 730000, China}

\begin{abstract}
Based on the isospin-dependent Boltzmann-Uehling-Uhlenbeck
transport model plus the Skyrme force parameters, initialization
effect is studied in heavy-ion collision at intermediate energies.
We find that there are moderate initialization effects in the
observables of free neutron to proton ratio ($n/p$), $\pi^{-}/\pi%
^{+}$ ratio, as well as neutron to proton differential flow
($F^{x}_{n-p}$). Effects of initialization are larger for charged
$\pi^{-}/\pi ^{+}$ ratios than $n/p$ ratios. And the effects of
initialization are more evident in nuclear reactions at lower
incident beam energies. We do not see large effects of
initialization for light reaction systems or large asymmetric
(neutron-richer) reaction systems. We also see relatively large
effects of initialization on the neutron to proton differential
flow at relatively lower incident beam energies or with large
impact parameters. These results may be useful for the delicate
studies of Equation of Sate (EoS) of asymmetric nuclear matter.
\end{abstract}

 \maketitle

\section{Introduction}

The Equation of State (EoS) of asymmetric nuclear matter becomes
one of the hot topics in today's nuclear physics, simply because
its isovector part, i.e., the symmetry energy, is a fundamental
and crucial ingredient in the investigations of exotic nuclei,
heavy-ion collisions, and astrophysical phenomena
\cite{natowitz10,xiao09,ditoro1,LCK08,Sum94,Lat04,Ste05a}. While
currently the theoretical predictions on the symmetry energy are
quite diverse
\cite{Che07,LiZH06,Pan72,Fri81,Wir88a,Kra06,Szm06,Bro00,Cha97,Sto03,Che05b,Dec80,MS,Kho96,Bas07,Ban00}.
Heavy-ion reactions induced by neutron-rich nuclei, especially
radioactive beams, provide a unique opportunity to constrain the
symmetry energy term in the EOS of isospin asymmetric nuclear
matter
\cite{ireview,ibook,baran05,tsang09,shetty07,fami06,tsang04,chen05}.
Nuclear reaction induced by the proton-rich ($N<Z$) nuclei was
also studied recently \cite{yong102}. Of particular interest is to
identify experimental observables that are sensitive to the
density dependence of nuclear symmetry energy. However, it is very
challenging to find such observables since the compression phase
is formed only transiently in heavy-ion reactions. Moreover, most
hadronic observables are affected by both the isospin symmetric
and asymmetric parts of the EOS at all densities throughout the
whole dynamical evolution of the reaction. And experimental
observables are also affected by nucleon-nucleon ($NN$) cross
sections \cite{yong101} and momentum dependence of isovector
mean-field potential \cite{gao10}. Thus rather delicate
observables have to be selected to probe the density dependence of
the nuclear symmetry energy, especially at high densities. The
neutron to proton ratio \cite{yong105,lcyz,lyz,ylc,tsang09,ma10},
the neutron-proton bremsstrahlung \cite{ylc08}, the $\pi
^{-}/\pi^{+}$ ratio
\cite{ba02a,gai04,qli05a,qli05b,yong06,xiao09}, the neutron-proton
differential flow \cite{li00,yong062} and the $K^{0}/K^{+}$ ratio
\cite{ditoro06} were proposed for this purpose. In all the above
studies, initialization effect in heavy-ion collisions at
intermediate energies was seldom mentioned. In this note, based on
the isospin-dependent Boltzmann-Uehling-Uhlenbeck transport model
plus the Skyrme force parameters, we studied initialization effect
in heavy-ion collisions at intermediate energies. We find that
there are moderate initialization effects in some common
experimental observables relating to nuclear symmetry energy.

\section{The theoretical methods}

The isospin and momentum-dependent mean-field potential used in
the present work is \cite{Das03}
\begin{eqnarray}
U(\rho, \delta, \textbf{p},\tau)
=A_u(x)\frac{\rho_{\tau^\prime}}{\rho_0}+A_l(x)\frac{\rho_{\tau}}{\rho_0}\nonumber\\
+B\left(\frac{\rho}{\rho_0}\right)^\sigma\left(1-x\delta^2\right)\nonumber
-8x\tau\frac{B}{\sigma+1}\frac{\rho^{\sigma-1}}{\rho_0^\sigma}\delta\rho_{\tau^{\prime}}\nonumber\\
+\sum_{t=\tau,\tau^{\prime}}\frac{2C_{\tau,t}}{\rho_0}\int{d^3\textbf{p}^{\prime}\frac{f_{t}(\textbf{r},
\textbf{p}^{\prime})}{1+\left(\textbf{p}-
\textbf{p}^{\prime}\right)^2/\Lambda^2}},
\label{Un}
\end{eqnarray}
where $\rho_n$ and $\rho_p$ denote neutron ($\tau=1/2$) and proton
($\tau=-1/2$) densities, respectively.
$\delta=(\rho_n-\rho_p)/(\rho_n+\rho_p)$ is the isospin asymmetry
of nuclear medium. All parameters in the preceding equation can be
found in refs. \cite{IBUU04}. The variable $x$ is introduced to
mimic different forms of the symmetry energy predicted by various
many-body theories without changing any property of symmetric
nuclear matter and the value of symmetry energy at normal density
$\rho_0$. Because the purpose of present studies is just to see
how large the effect of nuclear initialization on experimental
observables, we just let the variable $x$ be $0$. The main
characteristic of the present single particle is the momentum
dependence of nuclear symmetry potential, which has evident effect
on energetic free $n/p$ ratio and $\pi^{-}/\pi ^{+}$ ratio in
heavy-ion collisions \cite{IBUU04,gao10}. In the present work, we
use the isospin-dependent in-medium reduced $NN$ elastic
scattering cross section from the scaling model according to
nucleon effective mass \cite{factor,neg,pan,gale}. For in-medium
$NN$ inelastic scattering cross section, we use the forms in free
space since it is quite controversial. For nuclear
initializations, we use the Skyrme-Hartree-Fock with Skyrme
$M^{*}$ (SM) and Skyrme $1$ (S1) force parameters
\cite{Friedrich86}.

\begin{figure}[tbh]
\begin{center}
\includegraphics[width=0.5\textwidth]{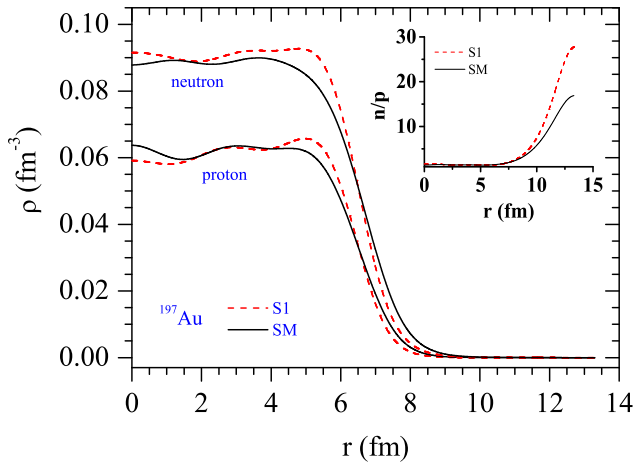}
\end{center}
\caption{(Color online) Density distributions of protons and
neutrons of nucleus $^{197}$Au with Skyrme $M^{*}$ (SM) and Skyrme
$1$ (S1) force parameters \cite{Friedrich86}, respectively. The
inset figure shows $n/p$ of nucleus $^{197}$Au as a function of
radius.} \label{dis}
\end{figure}
Before studying initialization effect in heavy-ion collisions, we
first give nucleonic density distributions of a nucleus $^{197}$Au
with Skyrme $M^{*}$ (SM) and Skyrme $1$ (S1) force parameters,
respectively. From Fig.~\ref{dis} we can see that nucleonic
distributions are more diffused with the SM force parameters. More
clearly, from the inset figure of Fig.~\ref{dis}, we can see that
$n/p$ of nucleonic distributions splits with the two set of force
parameters in the marginal area of nucleus while the values of
$n/p$ are almost the same in the core of nucleus. The SM
parameters match small $n/p$ while the S1 parameters give large
$n/p$. We also initialized the light nuclei $^{40}$Ca and
$^{48}$Ca, their nucleonic distributions show the same features.

\section{Results and discussions}

\begin{figure}[tbh]
\begin{center}
\includegraphics[width=0.5\textwidth]{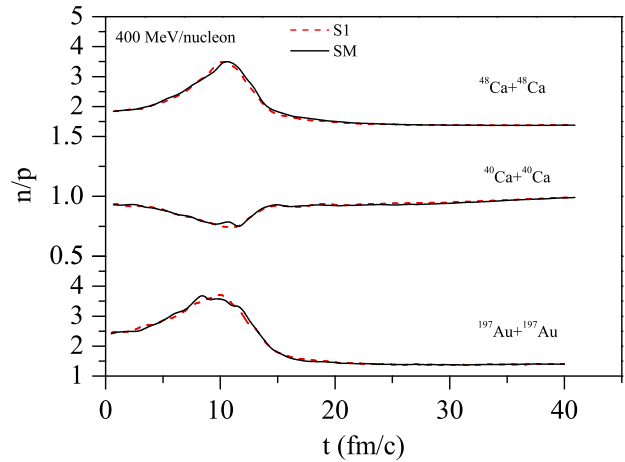}
\end{center}
\caption{(Color online) Effects of initialization of free neutron
to proton ratios of preequilibrium nucleon emissions in central
reactions of $^{48}$Ca+$^{48}$Ca, $^{40}$Ca+$^{40}$Ca and
$^{197}$Au+$^{197}$Au at 400 MeV/nucleon.} \label{rnp}
\end{figure}
We first see effects of initialization on free $n/p$ since the
free $n/p$ is a very common probe of nuclear symmetry energy
\cite{yong105,lcyz,lyz,ylc,tsang09,ma10}. Fig.~\ref{rnp} shows
initialization effect on the free $n/p$. We can see that for all
the reaction systems, whether light or heavy, there are almost no
initialization effects. Before this, one always think that effects
of initialization are large for light reaction system. Our studies
show that we can thoroughly neglect effects of initialization on
free $n/p$ of preeqilibrium emission. From Fig.~\ref{rnp}, we can
also see that effects of initialization on free $n/p$ are not
affected by asymmetry of reaction system.

\begin{figure}[tbh]
\begin{center}
\includegraphics[width=0.5\textwidth]{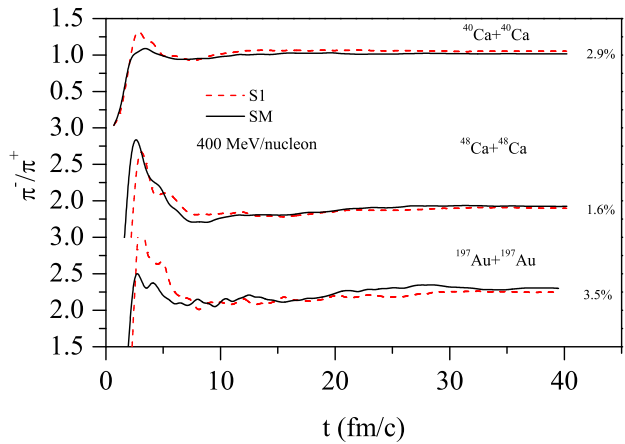}
\end{center}
\caption{(Color online) Effects of initialization on charged
$\pi^{-}/\pi ^{+}$ ratio in central reactions of
$^{48}$Ca+$^{48}$Ca, $^{40}$Ca+$^{40}$Ca and $^{197}$Au+$^{197}$Au
at 400 MeV/nucleon.} \label{rpion}
\end{figure}
We next turn to the studies of effects of initialization on
$\pi^{-}/\pi ^{+}$ ratio. $\pi^{-}/\pi ^{+}$ ratio recently
becomes a very hot probe owing to its connection with the
high-density behavior of nuclear symmetry energy
\cite{ba02a,gai04,qli05a,qli05b,yong06,xiao09}. Fig.~\ref{rpion}
shows effects of initialization on charged $\pi^{-}/\pi ^{+}$
ratio in central reactions of $^{48}$Ca+$^{48}$Ca,
$^{40}$Ca+$^{40}$Ca and $^{197}$Au+$^{197}$Au at 400 MeV/nucleon.
We can first see that effects of initialization are larger on
$\pi^{-}/\pi ^{+}$ ratio than free $n/p$. For heavy system
$^{197}$Au+$^{197}$Au, effects of initialization are larger than
light systems $^{48}$Ca+$^{48}$Ca and $^{40}$Ca+$^{40}$Ca. Again
we here do not see a large initialization effect for light system.
And we also do not see a large initialization effect on
$\pi^{-}/\pi ^{+}$ ratio from the reaction system with larger
asymmetry. For S1 force parameters, we see small $\pi^{-}/\pi
^{+}$ ratio than SM force parameters for the two large N/Z systems
$^{197}$Au+$^{197}$Au and $^{48}$Ca+$^{48}$Ca. This is because the
S1 force parameters make the N/Z of nucleus in marginal area large
(as shown in Fig.~\ref{dis}). The central area of nucleus, which
goes though a large compression and produces pions, matches small
N/Z. While the $\pi ^{-}/\pi ^{+}$ ratio is approximately
$(5N^{2}+NZ)/(5Z^{2}+NZ)\approx (N/Z)^{2}$ in central heavy-ion
reactions with $N$ and $Z$ being the total neutron and proton
numbers in the participant region \cite{stock},
S1 force parameters thus matches small values of $\pi^{-}/\pi%
^{+}$ ratio. In the calculations, effects of initialization on the
central reactions of $^{48}$Ca+$^{48}$Ca, $^{40}$Ca+$^{40}$Ca and
$^{197}$Au+$^{197}$Au at 400 MeV/nucleon are 1.6\%, 2.9\% and
3.5\%, respectively. We therefore can almost neglect
initialization effect in heavy-ion collisions while using charged
pion ratio to probe the density-dependent symmetry energy.

\begin{figure}[tbh]
\begin{center}
\includegraphics[width=0.5\textwidth]{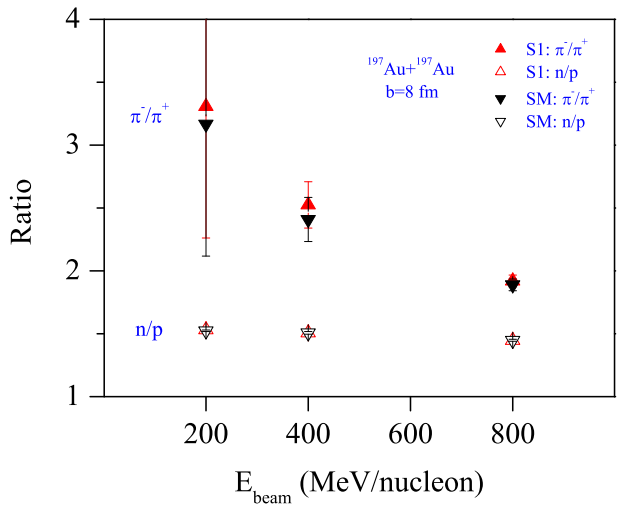}
\end{center}
\caption{(Color online) Beam energy dependence of initialization
effects on charged $\pi^{-}/\pi ^{+}$ ratio and free $n/p$ ratio
in the $^{197}$Au+$^{197}$Au reaction with an impact parameter of
8 fm.} \label{edep}
\end{figure}
Studies show that charged $\pi^{-}/\pi ^{+}$ ratio may be more
sensitive to nuclear symmetry energy at lower incident beam
energies via subthreshold pion production \cite{xiao09}. We thus
make a study of beam energy dependence of initialization effects
on charged $\pi^{-}/\pi ^{+}$ ratio as well as free $n/p$ ratio in
the $^{197}$Au+$^{197}$Au reaction with an impact parameter of 8
fm. From Fig.~\ref{edep}, we again see that free $n/p$ of
preequilibrium emission shows no initialization effects with the
decrease of beam energy. While for $\pi^{-}/\pi ^{+}$ ratio,
effects of initialization seems larger for lower incident beam
energies. Also we can clearly see that the values of $\pi^{-}/\pi
^{+}$ ratio become larger with the decrease of beam energy. This
trend is consistent with recent studies \cite{xiao09}. Comparing
with the case of SM, S1 always gives large $\pi^{-}/\pi ^{+}$
ratio in semi-central collision owing to the large N/Z of
compression phase (as shown in Fig.~\ref{dis}). Therefore, it
seems necessary to consider initialization effect of nucleus while
using $\pi^{-}/\pi ^{+}$ ratio to study nuclear symmetry energy
via subthreshold pion production. One way is that the force
parameters used for the initialization of nucleus give rough the
``correct'' symmetry energy value around saturation density. The
other way is making the force parameters give the same symmetry
energy, which can be deduced by the mean-field potential used in
the model.

\begin{figure}[tbh]
\begin{center}
\includegraphics[width=0.5\textwidth]{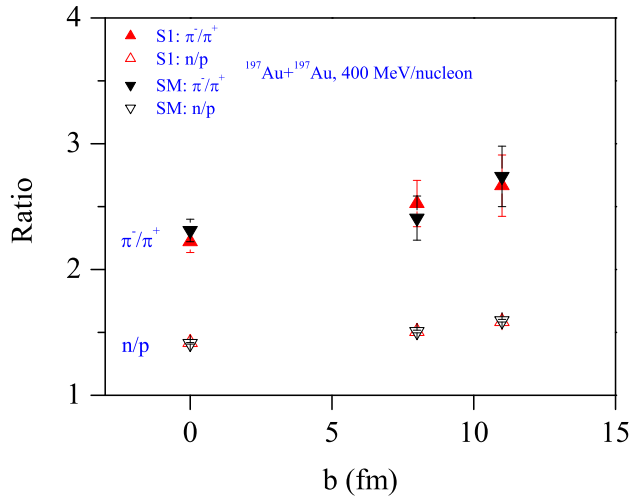}
\end{center}
\caption{(Color online) Impact parameter dependence of
initialization effects on charged $\pi^{-}/\pi ^{+}$ ratio and
free $n/p$ ratio in the $^{197}$Au+$^{197}$Au reaction at 400
MeV/nucleon.} \label{bdep}
\end{figure}
It is instructive to make a study of impact parameter dependence
of charged pion ratio in heavy-ion collisions at intermediate
energies. Fig.~\ref{bdep} shows impact parameter dependence of
initialization effects on charged $\pi^{-}/\pi ^{+}$ ratio and
free $n/p$ ratio in the $^{197}$Au+$^{197}$Au reaction at 400
MeV/nucleon. We can first see that the values of $\pi^{-}/\pi
^{+}$ ratio increase with the impact parameter owing to the large
N/Z of compression phase. Initialization effect, however, does not
keep stable. Comparing with the case of SM, S1 gives small
$\pi^{-}/\pi ^{+}$ ratio in central collision and shows large
$\pi^{-}/\pi ^{+}$ ratio in semi-central collision and for
marginal collision S1 again provides small $\pi^{-}/\pi ^{+}$
ratio. We also see that free $n/p$ of preequilibrium emission
shows no initialization effects with the increase of impact
parameter. We thus conclude that impact parameter dependence of
initialization effects on charged $\pi^{-}/\pi ^{+}$ ratio can be
neglected in heavy-ion collisions at intermediate energies.

\begin{figure}[tbh]
\begin{center}
\includegraphics[width=0.5\textwidth]{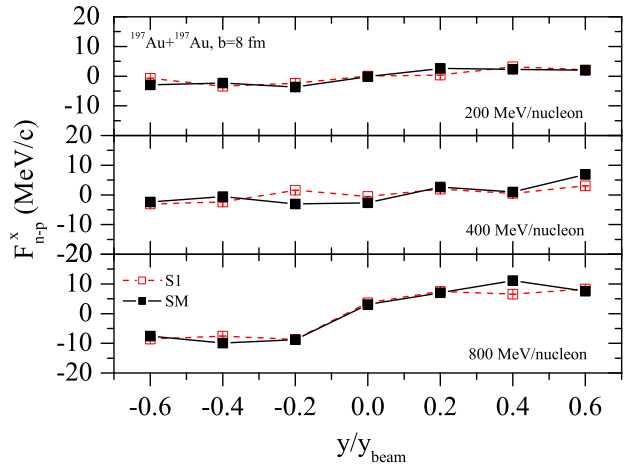}
\end{center}
\caption{(Color online) Effects of initialization of
neutron-proton differential flow (as a function of reduced
rapidity in C.M.) in $^{197}$Au+$^{197}$Au reaction at beam
energies of 200, 400 and 800 MeV/nucleon, respectively.}
\label{flow}
\end{figure}
\begin{figure}[tbh]
\begin{center}
\includegraphics[width=0.5\textwidth]{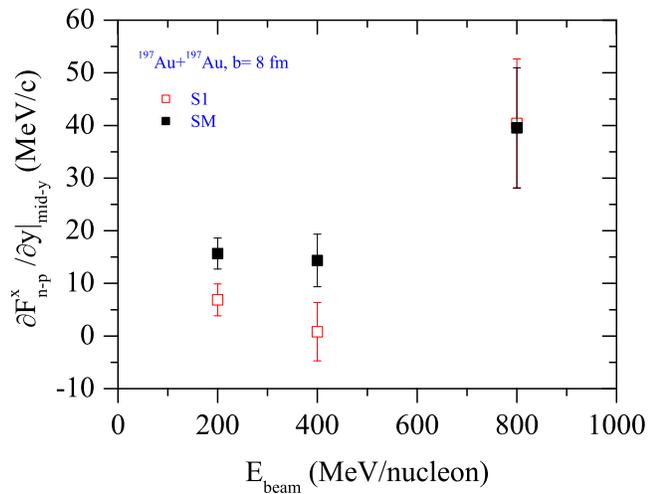}
\end{center}
\caption{(Color online) Beam energy dependence of the effects of
initialization of neutron-proton differential flow parameter
(slope of $F^{x}_{n-p}$ at mid-rapidities) in
$^{197}$Au+$^{197}$Au reaction with an impact parameter of 8 fm.}
\label{eflow}
\end{figure}
At last we turn to the studies of neutron-proton differential flow
\cite{li00,yong062}. The neutron-proton differential transverse
flow was defined as \cite{yong062}
\begin{eqnarray}
F_{n-p}^{x}(y) &\equiv
&\frac{1}{N(y)}\sum_{i=1}^{N(y)}p_{i}^{x}(y)w_{i}
\nonumber \\
&=&\frac{N_{n}(y)}{N(y)}\langle p_{n}^{x}(y)\rangle -\frac{N_{p}(y)}{N(y)}%
\langle p_{p}^{x}(y)\rangle  \label{npflow}
\end{eqnarray}%
where $N(y)$, $N_{n}(y)$ and $N_{p}(y)$ are the number of free
nucleons, neutrons and protons, respectively, at rapidity $y$;
$p_{i}^{x}(y)$ is the transverse momentum of the free nucleon at
rapidity $y$; $w_{i}=1$ $(-1)$ for neutrons (protons); and
$\langle p_{n}^{x}(y)\rangle $ and $\langle p_{p}^{x}(y)\rangle $
are respectively the average transverse momenta of neutrons and
protons at rapidity $y$. The neutron-proton differential flow
combines effects due to both the isospin fractionation and the
different transverse flows of neutrons and protons.
Fig.~\ref{flow} shows the neutron-proton differential flow as a
function of reduced rapidity in the framework of Center of Mass.
It is seen that effects of initialization on neutron-proton
differential flow are large at lower incident beam energies.
Initialization effect disappears at relatively large incident beam
energies. Fig.~\ref{eflow} clearly show that we should consider
the effects of nucleus initialization at relatively lower incident
beam energies. We can also see that flow parameter is large with
the SM force parameters (which makes the nucleonic distributions
more diffused as shown in Fig.~\ref{dis}).

\begin{figure}[tbh]
\begin{center}
\includegraphics[width=0.5\textwidth]{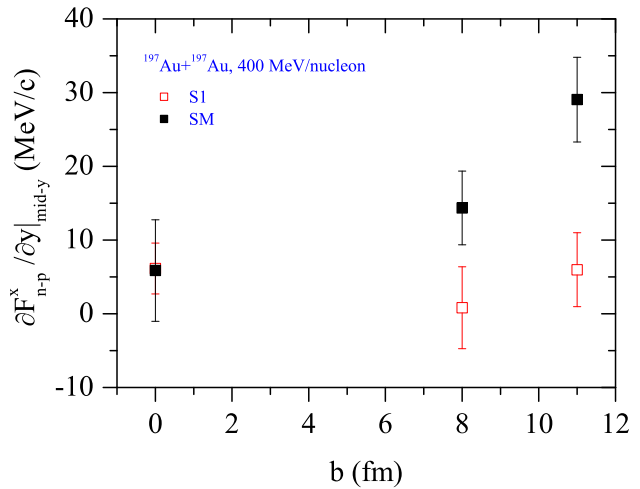}
\end{center}
\caption{(Color online) Impact parameter dependence of the effects
of initialization of neutron-proton differential flow parameter in
$^{197}$Au+$^{197}$Au reaction at beam energy of 400 MeV/nucleon.}
\label{bflow}
\end{figure}
Fig.~\ref{bflow} shows impact parameter dependence of the effects
of initialization on neutron-proton differential flow parameter.
It is clearly shown that effects of initialization on
neutron-proton differential flow parameter are large with the
increase of impact parameter owing to large difference of
nucleonic flow of emitting nucleons.

\section{Conclusions}

In conclusion, based on a transport model IBUU plus the Skyrme
force parameters, we studied the effects of initialization on
several probes in heavy-ion collisions at intermediate energies.
We find that for particle emission, reactions at lower incident
beam energies show relatively large initialization effects. And we
do not see large initialization effects for lighter reaction
systems. We also see large initialization effects on
neutron-proton differential flow at lower incident beam energies
or with large impact parameters. These results may be useful for
the transport model simulations. And they may be also useful for
the delicate studies of Equation of Sate of asymmetric nuclear
matter.

\section*{Acknowledgments}

The author G.C. Yong thanks Prof. Bao-An Li and Prof. Lie-Wen Chen
for useful discussions. The work is supported by the National
Natural Science Foundation of China (10875151, 10740420550), the
Knowledge Innovation Project (KJCX2-EW-N01) of Chinese Academy of
Sciences, the Major State Basic Research Developing Program of
China under No. 2007CB815004, the CAS/SAFEA International
Partnership Program for Creative Research Teams (CXTD-J2005-1) and
the Zhejiang Provincial Natural Science Foundation of China (Grant
No. Y6110644)

\end{document}